\documentclass[aps,twocolumn,prd,showpacs,nofootinbib]{revtex4}
\usepackage{amsmath} \usepackage{graphicx} \usepackage{dcolumn}
\usepackage{amssymb}

\newcommand{\si}[1]{{\scriptscriptstyle{#1}}}

\newcommand{\ud}{\mathrm{d}}
\newcommand{\ue}{\mathrm{e}}
\newcommand{\uh}{\mathrm{h}}
\newcommand{\ub}{\mathrm{b}}
\newcommand{\uJ}{\mathrm{J}}
\newcommand{\uY}{\mathrm{Y}}
\newcommand{\uI}{\mathrm{I}}
\newcommand{\uK}{\mathrm{K}}

\newcommand{\calD}{\mathcal{D}}

\newcommand{\matter}{\mathrm{mat}}

\newcommand{\ie}{i.e.~}
\newcommand{\eg}{e.g.~}

\newcommand{\rdim}{r}
\newcommand{\radim}{\rho}
\newcommand{\sqrgaadim}{\omega}
\newcommand{\sqrgaa}{\varpi}
\newcommand{\expsig}{\gamma}

\newcommand{\calL}{\mathcal{L}}
\newcommand{\calF}{\mathcal{F}}

\newcommand{\calE}{\mathcal{E}}
\newcommand{\calT}{\mathcal{T}}

\newcommand{\calP}{\mathcal{P}}

\newcommand{\calA}{\mathcal{A}}
\newcommand{\besselJ}[1]{\uJ{_{#1}}}
\newcommand{\besselY}[1]{\uY{_{#1}}}
\newcommand{\besselI}[1]{\uI{_{#1}}}
\newcommand{\besselK}[1]{\uK{_{#1}}}

\newcommand{\Omegan}{\Omega_n}
\newcommand{\harm}{\Upsilon}

\newcommand{\Mpbulk}{M_*}
\newcommand{\Mp}{M_{\mathrm{p}}}
\newcommand{\Hubble}{H}
\newcommand{\eV}{\mathrm{eV}}

\newcommand{\ug}{\mathrm{g}}
\newcommand{\um}{\mathrm{m}}

\newcommand{\uc}{\mathrm{c}}

\newcommand{\kappasix}{\kappa_{_6}}
\newcommand{\kappan}{\kappa_{_\ntot}}

\newcommand{\Sdis}{\tilde{S}}
\newcommand{\Sgdis}{\tilde{S}_{\ug}}
\newcommand{\Smdis}{\tilde{S}_{\um}}

\newcommand{\Hg}{H}

\newcommand{\mdil}{m_\ud}
\newcommand{\mhig}{m_\uh}
\newcommand{\mbos}{m_\ub}
\newcommand{\mdim}{M}
\newcommand{\madim}{\bar{\mdim}}

\newcommand{\Deltafunc}{\Delta}

\newcommand{\tens}{h}

\newcommand{\tensrescal}{\xi}
\newcommand{\tensspec}{\hat{\tens}}
\newcommand{\Boxadim}{\tilde{\Box}}
\newcommand{\Vtens}[1]{V_{#1}}
\newcommand{\Vnewt}{V_\ug}
\newcommand{\Uhat}{\hat{U}}

\newcommand{\ntot}{N}
\newcommand{\nextra}{\ntot_{\uc}}

\newcommand{\dgp}{\textrm{DGP}}

\newcommand{\dec}{\textrm{DEC}}
\newcommand{\wkb}{\textrm{WKB}}
\newcommand{\vev}{\textrm{VEV}}

\newcommand{\curv}{{\mathit{k}}}
\newcommand{\angmom}{\mathrm{L}}

\begin{document}

\title{Metastable gravity on classical defects}

\author{Christophe Ringeval}
\email{c.ringeval@imperial.ac.uk}
\affiliation{Blackett Laboratory, Imperial College, Prince 
Consort Road, London SW7 2AZ, United Kingdom}

\author{Jan-Willem Rombouts} \email{jwr218@nyu.edu}
\affiliation{Department of Physics, New York University, 4 
Washington Place, New York, NY 10003, USA}

\date{\today}

\begin{abstract}

We discuss the realization of metastable gravity on classical defects
in infinite-volume extra dimensions. In dilatonic Einstein gravity, it
is found that the existence of metastable gravity on the defect core
requires violation of the Dominant Energy Condition for codimension
$\nextra=2$ defects.  This is illustrated with a detailed analysis of
a six-dimensional hyperstring minimally coupled to dilaton gravity. We
present the general conditions under which a codimension $\nextra > 2$
defect admits metastable modes, and find that they differ from lower
codimensional models in that, under certain conditions, they do not
require violation of energy conditions to support quasi-localized
gravity.

\end{abstract}
\maketitle

\section{Introduction}
\label{sec:intro}
In recent years, there has been great interest in models of
infinite-volume extra dimensions. The Dvali-Gabadadze-Porrati (\dgp{})
model realizes four-dimensional ($4D$) gravity on a membrane in a flat
five-dimensional ($5D$) bulk by adding a large, induced
Einstein-Hilbert term to the worldvolume of a
three-brane~\cite{Dvali:2000hr}. The Einstein-Hilbert term on the
brane is generically induced by quantum loops of matter fields on the
brane, and realizations of this mechanism are known in more
fundamental theories like string
theory~\cite{Antoniadis:2002tr,Kohlprath:2003pu,
Kohlprath:2004yw}. One of the main phenomenological attractions of
\dgp{} is its celebrated ``self-accelerated'' cosmological solution,
giving small late-time acceleration without the need to invoke dark
energy~\cite{Deffayet:2000uy}.  Higher codimensional realizations of
\dgp{} gravity are known to be non-trivial, because a straightforward
generalization of the model gives singular propagation and other
difficulties~\cite{Dvali:2000xg,Dvali:2001ae,Dvali:2002pe,Gabadadze:2004dq,
Gabadadze:2003ck,Kolanovic:2003am}.

On the other hand, codimension $\nextra>1$ theories have very
attractive properties, foremost because of their unique way of
addressing the cosmological constant
problem~\cite{Dvali:2002pe,Gabadadze:2004dq}.  Being non-local in the
infrared, these models evade Weinberg's no-go theorem, unlike compact
extra dimensional theories. The idea put forward in
\cite{Dvali:2002pe} for example is that, in infinite-volume extra
dimensional theories, one can have $4D$ nearly flat solutions even
though the brane tension is large (``the $4D$ cosmological constant
curves the bulk, not the brane''). In more detail, using the induced
gravity setting, it was argued that there are solutions of
supercritical branes in infinite bulk with tension $\calT>\Mpbulk^4$
that support a regular geometry if one allows for an inflating
worldvolume with Hubble rate:
\begin{equation}
\Hubble \sim \Mpbulk \left(\dfrac{\Mpbulk^4}{\calT}\right)^{1/(\nextra-2)}.
\end{equation}
$\Mpbulk$ is the fundamental higher dimensional Planck mass
(constrained to be quite low $\sim 10^{-3} \eV$) and $\nextra$ the
number of codimensions. In predicting an inflation rate inversely
proportional to the vacuum energy density these models provide an
interesting way of solving the cosmological constant problem.

In Refs.~\cite{Gabadadze:2003ck,Kolanovic:2003am, Kolanovic:2003da,
Porrati:2004yi} different ways of realizing \dgp{}-like gravity in
higher dimensional context were discussed. A fundamental difference
with the codimension one model is that $4D$ gravity is mediated there by
massive resonant graviton states, with the width of the resonance much
smaller than its mass~\cite{Kolanovic:2003am} (\dgp{} allows $4D$
gravity by a broad resonance peaked around zero mass). However, the
general task remains to find a physical framework where gravitational
dynamics can be approximated by a \dgp-like action
\begin{equation}
\label{eq:action_dgp}
S=\Mp^2\int{\sqrt{|\bar{g}|} \, \bar{R} \,\ud^4 x} +
\Mpbulk^{2+\nextra} \int{ \sqrt{|g|} \,R \, \ud^{\nextra+4} X}.
\end{equation}
Here $\bar{g}$ and $\bar{R}$ are the determinant and scalar curvature
of the induced metric on the brane while $g$ and $R$ are the same
quantities for the bulk metric.
 
In analogy with the action~(\ref{eq:action_dgp}), in
Ref.~\cite{Kolanovic:2003am}, $4D$-like gravity is realized by a
``regularized'' action:
\begin{equation}
\label{eq:action_reg}
S= M_{*}^{2+\nextra}\int{
\sqrt{|g|} \,R \,{\cal F}(X) \, \ud^{\nextra+4} X},
\end{equation}
where ${\cal F}$ is a sharply peaked profile around the origin of the
transverse space, which can be interpreted as a varying Planck mass in
the transverse dimensions. This action has the same features as
Eq.~(\ref{eq:action_dgp}) in that gravity in the brane core is weakly
coupled while the bulk is strongly coupled. It was shown that by
appropriately choosing the profile ${\cal F}$, one propagates $4D$
gravity by resonances in the brane core\footnote{The varying
gravitational coupling constant defined by ${\cal F}$ is analogous to
a varying dielectric constant in electrostatics. The appearance of
resonant states is supported by this analogy.}. Indeed, the Newtonian
potential between localized sources on the brane, at a dimensionless
distance $x$ from eachother, can be obtained by considering the
exchange of Kaluza-Klein
modes~\cite{Randall:1999ee,Randall:1999vf,Garriga:1999yh}
\begin{equation}
\label{eq:vsketch}
\Vnewt(x) \propto \int_0^{\infty} \rho(\madim) \dfrac{\ue^{-\madim
x}}{x} \,\ud \madim.
\end{equation}
Here $\rho(\madim)$ is the spectral density on the brane, which is
modified from its empty space value in the presence of quasi-bound
states. For instance, in flat $5D$ spacetime, one recovers a static
potential varying as $1/x^2$ from a constant spectral density. One can
consider the case in which the spectral density is, in addition to its
standard behavior, strongly peaked around a particular mass
$\madim_\uc$
\begin{equation}
\rho(\madim) = 1 + \delta(\madim-\madim_\uc).
\end{equation}
{}From Eq.~(\ref{eq:vsketch}), this would lead to a gravitational
potential
\begin{equation}
\Vnewt(x) \propto \dfrac{1}{x} \left(\ue^{-\madim_\uc x} +
\dfrac{1}{x} \right),
\end{equation}
and one recovers $4D$-like gravity on the scales $1 \ll x \ll
1/\madim_\uc$, which is physically attractive for $\madim_\uc \ll
1$. Interestingly, such systems predict generically an infrared as
well as ultraviolet modification of gravity. In the regularization of
\cite{Kolanovic:2003am} this behaviour was indeed recovered, using
appropriate regulating profiles for the action
Eq.~(\ref{eq:action_reg}).

{}From Eq.~(\ref{eq:action_dgp}), it is clear that the previous
mechanism could also be realized by letting $\sqrt{|g|}$ vary
transverse to the brane instead of the Planck mass. This has been
studied in Refs.~\cite{Csaki:2000pp,Shaposhnikov:2004ds} in the
context of warped extra dimensions, with or without asymptotic
flatness.

Following the previous discussing, we want to focus in this paper on
the following question~\cite{Dvali:2000ty,Carroll:2001ih}: is it
possible to realize $4D$ gravity by (meta)stable states in the context
of classical field theories? Topological defects serve as natural
branes in non-perturbative field
theory~\cite{Akama:1982jy,Rubakov:1983bb,Visser:1985,Cvetic:1996vr,Carter:2002tk}
and they have been intensively studied in the context of warped extra
dimensions~\cite{Lee:2000cf,Bonjour:1999kz,Ringeval:2001cq,Peter:2003zg},
while an obvious setting to generate a varying Planck mass is
dilatonic
gravity~\cite{Brans:1961sx,Damour:1992we,Riazuelo:2001mg,Ellis:2003pw}. The
formulation of these theories using an underlying non-linear sigma
model realization is important to address their (supersymmetric)
phenomenology~\cite{Dvali:2000ty} and internal consistency (for debate
on the \dgp{} model see \eg Refs.~\cite{Luty:2003vm, Dvali:2004ph,
Nicolis:2004qq,Gabadadze:2004iy}).

Indeed, the action~(\ref{eq:action_reg}) suggests that in an
appropriately formulated theory of dilatonic gravity, one might
achieve metastable states in the core of a topological defect if the
dilaton condenses strongly around the core and falls off sharply
outside the defect. Clearly, such a model would involve both
variations in the metric through the defect energy-momentum tensor,
and variations of the ``bulk Planck mass'' by means of the dilaton. It
is not clear \emph{a priori} how these effects are balanced, and one
has to construct explicit solutions to make quantitative statements
about their properties.

This paper is organized as follows. We first construct an illustrative
example of a local hyperstring in six dimensional dilatonic
gravity. Considering a massive dilaton, we find that the dilaton
naturally condenses around the hyperstring core, in a region set by
its Compton wavelength, and that the geometry of the system is
asymptotically flat. The full non-linear solution is computed
numerically. Next we study the propagation of $4D$ tensor
perturbations on this background and find that there are no metastable
states appearing in this setting.

We then ask under which general conditions (on the defect forming
matter and gravity content) tensor (with respect to the $4D$ brane)
resonant states exist in the defect core in infinite-volume extra
dimensions. It will appear that a violation of the Dominant Energy
Condition (\dec{}) is a necessary condition for this to happen in
codimension $\nextra=2$.

In the last section we restate the same question for the case of
codimension $\nextra>2$ defects. It is shown that, at least under
reasonable conditions, metastable gravity is allowed without violation
of positive energy conditions. We leave however explicit constructions
for future work. 

Finally we conclude, and stress that our results only concern
classical field theoretical realizations of metastable gravity. As
mentioned before, there exist a number of different ways towards
realizing metastable gravity on branes in infinite-volume extra
dimensions, using the \dgp{} loop induced effects or explicit
constructions in string theory. Our work evidently does not address
those models.
    
\section{4D Hyperstring coupled to Dilatonic Gravity}

In the following we consider the action for a four-dimensional
hyperstring coupled to scalar-tensor gravity in a six-dimensional
spacetime~\cite{Damour:1992we,Brans:1961sx}:
\begin{equation}
\begin{aligned}
S&=\int \frac{1}{2\kappasix^2}\ue^\phi \left[ R - g^{\si{AB}}
\partial_{\si{A}} \phi \partial_{\si{B}} \phi - U(\phi)
\right]\sqrt{|g|} \,\ud^6 x  \\ &+\int
{\calL}_{\matter}\sqrt{|g|}\,\ud^6 x,
\end{aligned}
\label{eq:action}
\end{equation}
where $g_{\si{AB}}$ is the $6D$ metric with signature $(-,+,+,\dots)$,
$R$ its Ricci scalar, $\phi$ the dilaton field with a potential
$U(\phi)$, $\kappasix^2\equiv 32\pi^2G_{_6}/3$, $G_{_6}$ being the
$6$-dimensional gravity constant. In order to allow for topological
vortex configurations, we include a complex scalar field $\Phi$:
\begin{equation}\label{eq:lag}
{\calL}_{\matter} =
-\frac{1}{2}g^{\si{AB}}\left(\calD_{\si{A}}\Phi\right)^\dag
\calD_{\si{B}}\Phi -V(\Phi)-\frac{1}{4}\Hg_{\si{AB}}\Hg^{\si{AB}},
\end{equation}
in which capital Latin indexes $A,B\ldots$ run from 0 to 5 and
$\Hg_{\si{AB}}$ is the electromagnetic-like tensor defined by
\begin{equation}\label{eq:fab}
\Hg_{\si{AB}}=\partial_{\si{A}} C_{\si{B}} -\partial_{\si{B}}
C_{\si{A}},
\end{equation}
where $C_{\si{B}}$ is the 1-form connection. The U(1) covariant
derivative $\calD_{\si{A}}$ is defined by
\begin{equation}\label{eq:d}
\calD_{\si{A}}\equiv\partial_{\si{A}} -iqC_{\si{A}},
\end{equation}
where $q$ is the charge. The potential of the scalar field $\Phi$ is
chosen to break the underlying U(1) symmetry and thereby allow for
hyperstring configurations,
\begin{equation}\label{eq:V}
V(\Phi)=\frac{\lambda}{8}\left(\vert\Phi\vert^2-\eta^2\right)^2,
\end{equation}
where $\lambda$ is a coupling constant and $\eta=\langle
|\Phi|\rangle$ is the magnitude of the scalar field vacuum expectation
values (\vev{}).

The variations of the action (\ref{eq:action}) with respect to the
metric and the dilaton field lead to the equations of motion in the
scalar-tensor gravity sector, in the Jordan frame one gets
\begin{align}
\label{eq:tensor}
G_{\si{AB}} &= \ue^{-\phi} \kappasix^2 T_{\si{AB}} + 2 \partial_{\si{A}}
\phi \partial_{\si{B}} \phi - \dfrac{1}{2} g_{\si{AB}} \left[ 3
\partial_{\si{X}} \phi \partial^{\si{X}} \phi \right. \\ \nonumber &+
\left. U(\phi) \right] + \nabla_{\si{A}} \partial_{\si{B}} \phi -
g_{\si{AB}} \nabla_{\si{X}} \partial^{\si{X}} \phi, \\
\label{eq:scalar}
\nabla_{\si{X}}
\partial^{\si{X}} \phi & = \dfrac{1}{2} U(\phi) + \dfrac{1}{2}
\dfrac{\ud U}{\ud \phi} - \dfrac{1}{2} R - \dfrac{1}{2} \partial_{\si{X}} \phi
\partial^{\si{X}} \phi,
\end{align}
where $G_{\si{AB}}$ is the $\ntot$-dimensional Einstein tensor and
$T_{\si{AB}}$ the stress tensor of the hyperstring:
\begin{equation}
\label{eq:tmunugen}
T_{\si{AB}} \equiv -2\frac{\delta {\calL}_{\matter}}{\delta
g^{\si{AB}}} + g_{\si{AB}} {\calL}_{\matter}.
\end{equation}

The metric is chosen so as to respect the cylindrical static symmetry
in the two extra dimensions and Poincar\'e invariance along the
remaining four spacetime coordinates;
\begin{equation}\label{eq:metric}
\ud s^2=g_{\si{AB}} \ud x^{\si{A}} \ud x^{\si{B}} =
\ue^{\sigma(\rdim)}\eta_{\mu\nu} \ud x^\mu\ud x^\nu
+\ud\rdim^2+\sqrgaadim^2(\rdim) \ud\theta^2,
\end{equation}
where $\eta_{\mu\nu}$ is the four dimensional Minkowski metric of
signature ($-,+,+,+$), and $(\rdim,\theta)$ the polar coordinates in
the extra dimensions. Greek indexes $\mu,\nu\ldots$ run from 0 to 3
and describe the brane worldvolume.

We chose the Nielsen-Olesen ansatz for the matter field leading to an
hyperstring configuration with unit winding
number~\cite{Nielsen:1987fy,Peter:2003zg},
\begin{equation}\label{eq:ansatz}
\begin{aligned}
\Phi= \varphi(\rdim)\ue^{i \theta},& & C_\theta = \displaystyle
\frac{1}{q}\left[1-Q(\rdim)\right].
\end{aligned}
\end{equation}
The only non-vanishing component of the electromagnetic tensor is
$\Hg_{\theta\rdim}=Q'/q$. The dilaton is also assumed to respect the
cylindrical symmetry and depends only on the extra radial coordinates
$\phi(r)$.

\subsection{Background Solution}
\label{sec:motion}

In terms of the dimensionless coordinate
\begin{equation}
\label{eq:radim}
\radim \equiv \mhig \rdim = \sqrt{\lambda} \eta \, r,
\end{equation}
$\mhig$ being the mass of the Higgs boson, the Einstein tensor reads
\begin{equation}
\label{eq:gab}
\begin{aligned}
G_{\mu \nu} &= g_{\mu \nu} \mhig^2 \left(\dfrac{3}{2} \ddot{\sigma} +
\dfrac{3}{2} \dot{\sigma}^2 + \dfrac{\ddot{\sqrgaadim}}{\sqrgaadim} +
\dfrac{3}{2} \dot{\sigma} \dfrac{\dot{\sqrgaadim}}{\sqrgaadim}
\right),\\ G_{\rdim\rdim} &= \mhig^2 \left(\dfrac{3}{2} \dot{\sigma}^2
+ 2 \dot{\sigma} \dfrac{\dot{\sqrgaadim}}{\sqrgaadim} \right),\\
G_{\theta\theta} &= \sqrgaadim^2 \mhig^2 \left(2 \ddot{\sigma} +
\dfrac{5}{2} \dot{\sigma}^2 \right),
\end{aligned}
\end{equation}
where the dot stands for differentiation with respect to
$\radim$. {}From Eqs.~(\ref{eq:lag}) and (\ref{eq:tmunugen}), the stress
tensor reads
\begin{equation}
\label{eq:tab}
\begin{aligned}
T_{\mu \nu} &= -g_{\mu \nu} \left(\dfrac{1}{2} \mhig^2 \dot{\varphi}^2
+ \dfrac{1}{2} \dfrac{\varphi^2 Q^2}{ \sqrgaadim^2} + V(\varphi) +
\dfrac{1}{2} \dfrac{\mhig^2 \dot{Q}^2}{q^2 \sqrgaadim^2} \right),\\
T_{\rdim \rdim} &= \dfrac{1}{2}\mhig^2 \dot{\varphi}^2 - \dfrac{1}{2}
\dfrac{\varphi^2 Q^2}{\sqrgaadim^2} - V(\varphi) + \dfrac{1}{2}
\dfrac{\mhig^2 \dot{Q}^2}{q^2 \sqrgaadim^2}, \\ T_{\theta\theta} &=
\sqrgaadim^2 \left[ -\dfrac{1}{2}\mhig^2 \dot{\varphi}^2 +
\dfrac{1}{2} \dfrac{\varphi^2 Q^2}{\sqrgaadim^2} - V(\varphi) +
\dfrac{1}{2} \dfrac{\mhig^2 \dot{Q}^2}{q^2 \sqrgaadim^2} \right].
\end{aligned}
\end{equation}

In terms of dimensionless quantities, the Einstein and dilaton
equations of motion (\ref{eq:tensor}) and (\ref{eq:scalar}) can be
recast into
\begin{align}
\label{eq:Gmunu}
\dfrac{3}{2} \ddot{\sigma} + \dfrac{3}{2} \dot{\sigma}^2 &+
\dfrac{\ddot{\sqrgaa}}{\sqrgaa} + \dfrac{3}{2} \dot{\sigma}
\dfrac{\dot{\sqrgaa}}{\sqrgaa} = \alpha \ue^{-\phi} \Bigg[-\dot{f}^2
\nonumber - \dfrac{f^2 Q^2}{\sqrgaa^2} \\ &- \dfrac{1}{4}(f^2-1)^2 -
\dfrac{\dot{Q}^2}{\varepsilon \sqrgaa^2 } \Bigg] - \ddot{\phi} -
\dfrac{3}{2} \dot{\phi}^2 \nonumber \\ & - \left(\dfrac{3}{2}
\dot{\sigma} + \dfrac{\dot{\sqrgaa}}{\sqrgaa} \right)
\dot{\phi} - \dfrac{1}{2} \dfrac{U}{\mhig^2} ,
\end{align}
\begin{align}
\label{eq:Grr}
\dfrac{3}{2} \dot{\sigma}^2 +
2\dot{\sigma}\dfrac{\dot{\sqrgaa}}{\sqrgaa} &= \alpha \ue^{-\phi}
\left[\dot{f}^2 - \dfrac{f^2 Q^2}{\sqrgaa^2} - \dfrac{1}{4}(f^2-1)^2 +
\dfrac{\dot{Q}^2}{\varepsilon \sqrgaa^2 } \right] \nonumber \\ &+
\dfrac{1}{2} \dot{\phi}^2 - \left(2 \dot{\sigma} +
\dfrac{\dot{\sqrgaa}}{\sqrgaa} \right) \dot{\phi} - \dfrac{1}{2}
\dfrac{U}{\mhig^2},
\end{align}
\begin{align}
\label{eq:Gaa}
2 \ddot{\sigma} + \dfrac{5}{2} \dot{\sigma}^2 &= \alpha
\ue^{-\phi} \left[-\dot{f}^2 + \dfrac{f^2 Q^2}{\sqrgaa^2} -
\dfrac{1}{4}(f^2-1)^2 + \dfrac{\dot{Q}^2}{\varepsilon \sqrgaa^2 } \right]
\nonumber \\
&- \ddot{\phi} -\dfrac{3}{2} \dot{\phi}^2 - 2 \dot{\sigma} \dot{\phi} -
\dfrac{1}{2} \dfrac{U}{\mhig^2},
\end{align}
\begin{align}
\label{eq:dil}
\ddot{\phi} + \dfrac{1}{2}\dot{\phi}^2 &+ \left(2 \dot{\sigma} +
\dfrac{\dot{\sqrgaa}}{\sqrgaa} \right) \dot{\phi} =\dfrac{1}{2
\mhig^2} \left[U + \dfrac{\ud U}{\ud \phi} - R \right].
\end{align}
We have defined the dimensionless parameters
\begin{align}
\alpha \equiv \dfrac{1}{2} \kappasix^2 \eta^2 ,&& \beta\equiv
\dfrac{\mdil^2}{\mhig^2}, &&\varepsilon \equiv
\dfrac{\mbos^2}{\mhig^2}=\dfrac{q^2 \eta^2}{\lambda \eta^2},
\end{align}
where $\mbos$ is the mass of the gauge vector boson and $\mdil$ the
mass of the dilaton in the Einstein frame for
\begin{equation}
U(\phi)= \mdil^2 \phi^2 \ue^{\phi/2}.
\end{equation}
The Ricci scalar is given by
\begin{equation}
\label{eq:ricci}
R = -\mhig^2 \left(4 \ddot{\sigma} + 5 \dot{\sigma}^2 +
2\dfrac{\ddot{\sqrgaa}}{\sqrgaa} + 4 \dot{\sigma}
\dfrac{\dot{\sqrgaa}}{\sqrgaa} \right),
\end{equation}
and $\sqrgaa$ is the dimensionless angular metric factor
\begin{equation}
\label{eq:gaa}
\sqrgaa \equiv \mhig \sqrgaadim.
\end{equation}
The modulus of the Higgs field appears through the dimensionless
function $f(\radim)$
\begin{equation}
f \equiv \dfrac{\varphi}{\eta}.
\end{equation}
Note that the metric (\ref{eq:metric}) has no conical singularity in
$r=0$ provided $\sqrgaadim \sim r$, \ie $\sqrgaa \sim \radim$ in the
hyperstring core. Differentiation of the action (\ref{eq:action}) with
respect to the Higgs field yields the Klein-Gordon equation
\begin{equation}
\label{eq:kg}
\ddot f +\left(2\dot\sigma
+\dfrac{\dot{\sqrgaa}}{\sqrgaa}\right)\dot f -
\dfrac{f Q^2}{\sqrgaa^2} - \dfrac{1}{2} f \left(f^2-1\right) = 0,
\end{equation}
while differentiation with respect to the gauge field gives the
Maxwell equation
\begin{equation}
\label{eq:max}
\ddot Q + \left(2\dot\sigma-\dfrac{\dot{\sqrgaa}}{\sqrgaa}
\right)\dot Q - \varepsilon f^2 Q = 0.
\end{equation}
Since the dilaton is only coupled to the metric, the Bianchi
identities ensure that  stress-energy conservation is still
verified in the matter sector. Therefore, there is a redundant
equation in the system of differential equations (\ref{eq:Gmunu}) to
(\ref{eq:max}) and we choose to solve numerically
Eqs.~(\ref{eq:Gmunu}), (\ref{eq:Gaa}), (\ref{eq:dil}), (\ref{eq:kg})
and (\ref{eq:max}). 

\begin{figure}
\begin{center}
\includegraphics[width=8.5cm]{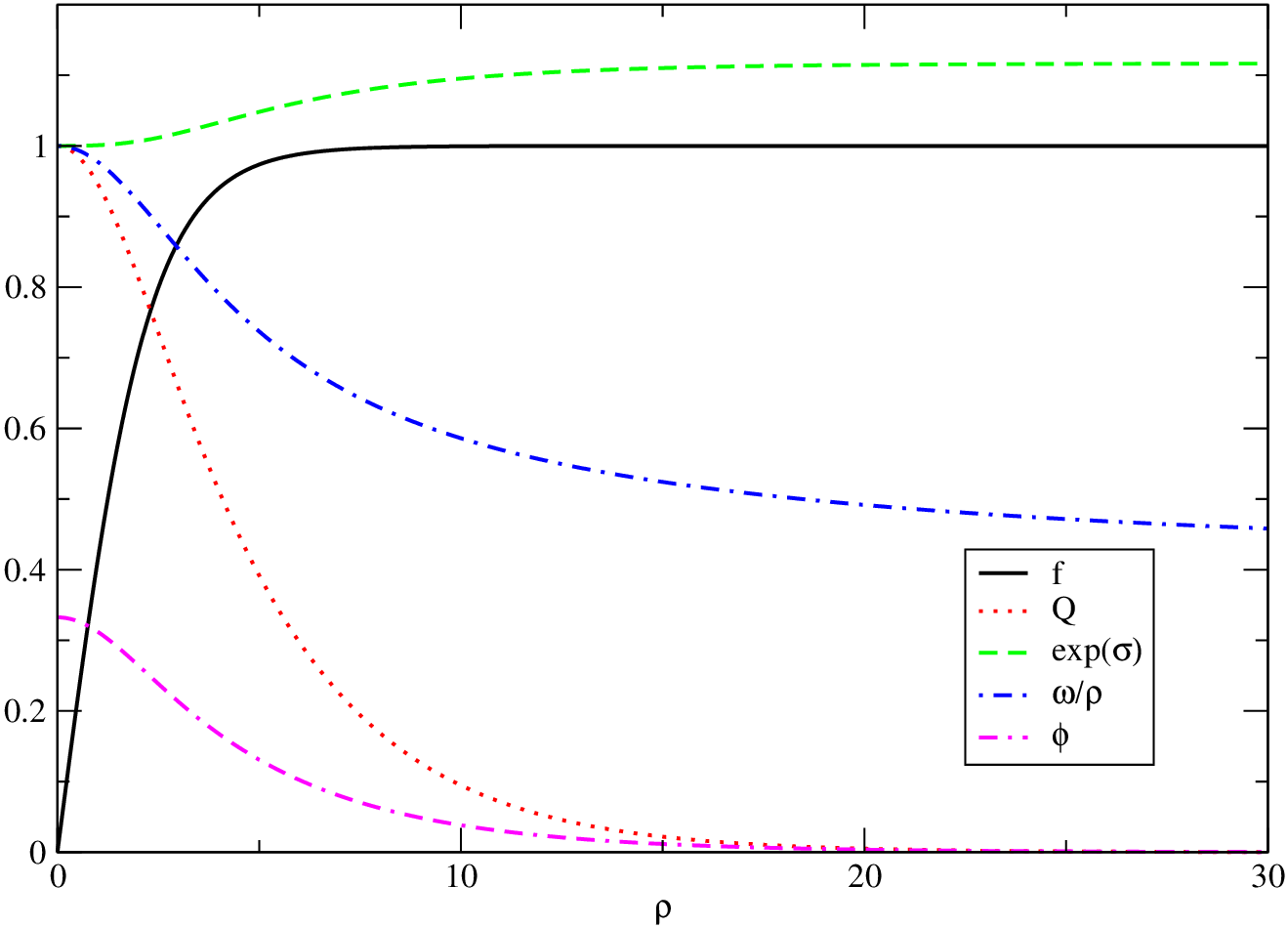}
\caption{Transverse profiles of the dimensionless hyperstring forming
Higgs $f$ and gauge field $Q$, together with the metric factors
$\ue^{\sigma}$ and $\sqrgaa/\radim$ for the parameters $\alpha=0.4$,
$\beta=0.1$ and $\varepsilon=0.1$. The dilaton $\phi$ condenses in the
hyperstring core and the spacetime has an asymptotically flat (conical)
geometry.}
\label{fig:dilvort}
\end{center}
\end{figure}

As usual, the Higgs field vanishes in the defect core, \ie $\Phi = 0$
for $\radim=0$, while it relaxes to its vacuum expectation value
(\vev{}), $\eta$, in the bulk. These conditions translate into the
following boundary conditions for the function $f$:
\begin{equation}
\label{eq:limhiggs}
f(0)=0,\qquad \lim_{\radim\to +\infty}f = 1.
\end{equation}
The corresponding boundary conditions for the gauge field are
given by
\begin{equation}
\label{eq:limgauge}
 Q(0)=1,\qquad \lim_{\radim\to +\infty}Q = 0.
\end{equation}
Moreover, we assume that no additional $\delta$-like energy
distribution lies in the hyperstring core. The geometry is therefore
regular in the core and the metric coefficients verify
\begin{equation}
\begin{aligned}
\sigma(0)=0, & & \dot{\sigma}(0)=0,\\
\sqrgaa(0)=0, & & \dot{\sqrgaa}(0)=1.
\end{aligned}
\end{equation}
Asymptotically we are interested in the cosmic string branch
solutions. Recall that the dilaton is massive and does not couple to
the matter sector allowing the existence of asymptotic flat conical
spacetimes without long range dilatonic
effects~~\cite{Gregory:1997wk,Verbin:1998tc,
Verbin:2001ye,deAndrade:2001mr}. This leads to the boundary conditions
\begin{equation}
\label{eq:limdil}
\dot{\phi}(0)=0,\qquad \lim_{\radim\to +\infty}\phi = 0,
\end{equation}
where the constraint at $\radim=0$ comes from the regularity
requirement of a non-winding field. The numerical method used to solve
this system is described in appendix~\ref{sect:appnum} and a typical
solution is presented in Fig.~\ref{fig:dilvort}.
The dilaton condenses in the hyperstring core while the spacetime
geometry is asymptotically conical, with a missing angle given by the
asymptotic value of $2 \pi(1-\sqrgaa/\radim)$ (see
Fig.~\ref{fig:dilvort}). Note also the red/blue shift $\exp(\sigma)$
between the core and the outer regions. In the next section, the
behaviour of the $4D$ tensor waves in this background is investigated,
both analytically and numerically, and we will find that this system
does not support quasi-localized gravity on the brane.

\subsection{Tensor Perturbations}
\label{sec:pert}

In this section, we investigate the propagation of tensor
perturbations, with respect to the $4D$ vortex worldvolume, on the
background field configuration studied in the previous section. For
this purpose we derive the equation of motion of tensor perturbations
in the Jordan metric, assuming that the scalar and vector degrees of
freedom are fixed.

Since we are interested in pure $4D$ tensor perturbations in the
Jordan frame, the perturbed metric (\ref{eq:metric}) reduces to
\begin{equation}
\label{eq:pertmetric}
\ud s^2 = \ue^{\sigma} \left( \eta_{\mu \nu} + \tens_{\mu \nu} \right)
\ud x^{\mu} \ud x^\nu + \dfrac{\ue^{\sigma}\ud z^2 + \sqrgaa^2 \ud
\theta^2}{\mhig^2},
\end{equation}
where $\tens_{\mu \nu}$ is a transverse and traceless tensor
\begin{equation}
\label{eq:dvtless}
\eta^{\mu \beta} \nabla_\beta \tens_{\mu \nu}=0, \qquad \eta^{\mu \nu}
\tens_{\mu \nu}=0,
\end{equation}
and $z$ is a dimensionless conformal coordinate defined by
\begin{equation}
\label{eq:defz}
z \equiv \int_{0}^{\radim} \ue^{-\sigma(u)/2} \ud u.
\end{equation}

With the hyperstring forming fields being of the scalar kind for the
Higgs, and of the vector kind for the gauge boson, the tensor modes
decouple from the background fields. The perturbed equations of motion
are obtained by perturbing the Einstein-Jordan equations
(\ref{eq:tensor}) at first order. {}From the perturbed metric tensor
$\delta g_{\mu \nu}$ given in Eq.~(\ref{eq:pertmetric}), one obtains
in terms of the dimensionless quantities (see
Appendix~\ref{sect:apppert})
\begin{equation}
\label{eq:tensmotion}
\tens''_{\mu \nu} + \left(\dfrac{3}{2} \sigma' +
\dfrac{\sqrgaa'}{\sqrgaa} + \phi' \right) \tens'_{\mu \nu} + \Boxadim
\tens_{\mu \nu} + \dfrac{\ue^{\sigma}}{\sqrgaa^2}\partial_\theta^2
\tens_{\mu \nu} = 0,
\end{equation}
where the prime denotes the derivative with respect to $z$ and
$\Boxadim$ stands for the dimensionless four-dimensional d'Alembertian
\begin{equation}
\label{eq:boxadim}
\Boxadim \equiv \dfrac{\Box}{\mhig^2} = \dfrac{\eta^{\mu \nu}
  \partial_\mu \partial_\nu}{\mhig^2}.
\end{equation}

It is convenient to decompose these perturbations in terms of winding
modes around the hyperstring and d'Alembertian eigenfunctions:
\begin{equation}
\label{eq:tensmode}
\tens_{\alpha \beta}(x^{\mu},\radim,\theta) = \sum_{p}
\ue^{i p \theta} \int \tensspec_{\alpha
\beta}^{(p)}(\madim,\radim) \Deltafunc_{\madim}(x^\mu) \ud \madim,
\end{equation}
where $\Deltafunc_{\madim}$ verifies
\begin{equation}
\label{eq:tensspec}
\left(\Boxadim + \madim^2 \right) \Deltafunc_{\madim} = 0.
\end{equation}
Moreover, the rescaling
\begin{equation}
\label{eq:tensrescal}
\tensrescal_{\alpha \beta}^{(p)} = \ue^{3\sigma/4 + \phi/2}
\sqrt{\sqrgaa} \, \tensspec_{\alpha \beta}^{(p)},
\end{equation}
allows to recast Eq.~(\ref{eq:tensmotion}) in a Schr\"odinger-like form
for the $\tensrescal_{\alpha \beta}^{(p)}$ modes
\begin{equation}
\label{eq:schrod}
- \dfrac{\ud^2 \tensrescal}{\ud z^2} + \Vtens{p}(z) \tensrescal =
  \madim^2 \tensrescal,
\end{equation}
where the tensor and angular mode indexes have been removed. The
potential $\Vtens{p}$ is given by
\begin{equation}
\label{eq:pot}
\Vtens{p}=W^2 + W' + \dfrac{\ue^{\sigma}}{\sqrgaa^2} p^2,
\end{equation}
with
\begin{equation}
\label{eq:superpot}
W(z)=\dfrac{3}{4} \sigma' + \dfrac{1}{2} \dfrac{\sqrgaa'}{\sqrgaa} +
\dfrac{1}{2} \phi'.
\end{equation}
In Eq.~(\ref{eq:schrod}), $\madim^2$ stands for the dimensionless
d'Alembertian eigenvalue
\begin{equation}
\madim^2 \equiv \dfrac{\mdim^2}{\mhig^2} = -\dfrac{\eta^{\alpha \beta}
k_\alpha k_\beta}{\mhig^2}.
\end{equation}
Note that for the zero angular momentum modes $p=0$,
Eq.~(\ref{eq:schrod}) is supersymmetric and $W$ is the
superpotential~\cite{Cooper:1994eh,Giovannini:2001fh}. Indeed,
defining the operator
\begin{equation}
\calA \equiv \dfrac{\ud}{\ud z} + W(z),
\end{equation}
simplifies Eq.~(\ref{eq:schrod}) into
\begin{equation}
\left(\calA \calA^\dag + \dfrac{\ue^{\sigma}}{\sqrgaa^2} p^2 \right)
\tensrescal = \madim^2 \tensrescal.
\end{equation}
This ensures that all the eigenvalues $\madim^2$ are positive, and
thus there are no tachyons in this model, provided supersymmetry is
not broken, \ie there exists a normalizable zero mode.

\subsection{Quasi-Localized States and $4D$ Gravity}

\subsubsection{Pure Cylindrical Waves}

It is interesting to briefly recall the case of cylindrical waves
propagating in flat spacetime. In that case, without dilaton,
$\sigma=\phi=0$ and $\sqrgaa=z$, the potential (\ref{eq:pot})
simplifies into
\begin{equation}
\Vtens{p} = - \dfrac{1}{4 z^2} + \dfrac{p^2}{z^2},
\end{equation}
which is always negative definite for $p=0$ and positive definite
otherwise. Eq.~(\ref{eq:schrod}) is a Bessel equation whose regular
solutions in the hyperstring core are
\begin{equation}
\tensrescal^{(p)}(\madim,z) \propto \sqrt{z} \, \besselJ{p}(\madim z),
\end{equation}
the $\besselY{p}$ and $\besselK{p}$ modes being singular in $z=0$
whereas the $\besselI{p}$ modes are singular at infinity. Moreover,
since~\cite{Abramovitz:1970aa}
\begin{equation}
\label{eq:besselJsmall}
\besselJ{p}(z) \underset{0}\sim \dfrac{1}{\Gamma(p+1)}
\left(\dfrac{z}{2} \right)^p,
\end{equation}
only the zero angular momentum modes $p=0$ do not vanish in the string
core. Note that in the special case $\madim=0$ the regular solutions
simplify to $z^p$. Far from the hyperstring core
\begin{equation}
\tensrescal^{(p)}(\madim z) \underset{\infty}{\propto}
\sqrt{\dfrac{2}{\pi \madim}} \cos \negthinspace \left(\madim z - p \dfrac{\pi}{2}
-\dfrac{\pi}{4} \right),
\end{equation}
and using plane wave normalization at infinity, one
gets from Eq.~(\ref{eq:tensrescal}) the spectral density
\begin{equation}
\left|\tensspec^{(0)}(\madim,0)\right|^2 =
\dfrac{\pi}{2}\madim.
\end{equation}
These tensor modes therefore contribute to a Newton potential
on the brane~\cite{Kolanovic:2003am,Kolanovic:2003da}
\begin{equation}
\label{eq:vnewt}
\Vnewt(x) \propto \int_0^{\infty}
\left|\tensspec(\madim,0)\right|^2 \dfrac{\ue^{-\madim x}}{x} \,\ud \madim
\propto \dfrac{1}{x^3},
\end{equation}
\ie compatible with the standard Newtonian gravity in a
six-dimensional flat spacetime.

As discussed in the introduction, if there exist localized or
quasi-localized waves in the hyperstring the spectral density will be
modified, and hence so will the resulting potential in
Eq.~(\ref{eq:vnewt}).

\subsubsection{Gravitational Propagation on the Hyperstring Background}

\begin{figure}
\begin{center}
\includegraphics[width=8.5cm]{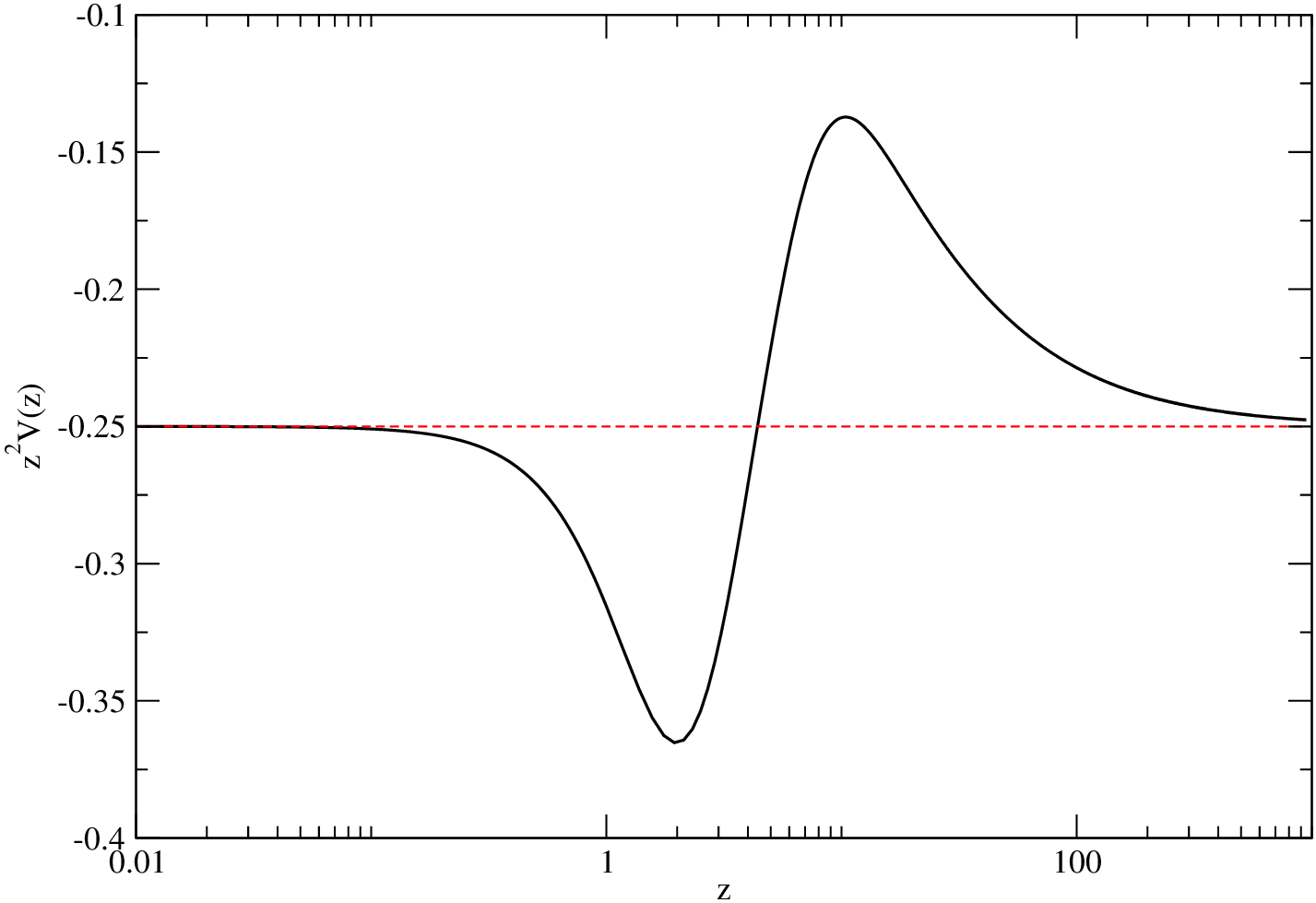}
\caption{The potential $z^2 \Vtens{0}(z)$ governing the behaviour of
  the tensor perturbations around the hyperstring according to
  Eq.~(\ref{eq:schrod}). The background fields are those of
  Fig.\ref{fig:dilvort} and the straight line represents the pure flat
  spacetime case.}
\label{fig:dilpot}
\end{center}
\end{figure}

According to the background solution computed in first section,
$\sigma'=\phi'=0$ and $\sqrgaa'/\sqrgaa=1/z$ both in the hyperstring
core and at infinity (see Fig.~\ref{fig:dilvort}). In these two
regions the potential $\Vtens{p}$ behaves like the flat spacetime one
and only the zero angular momentum modes $p=0$ can contribute
significantly to the spectral density. As can be seen in
Fig.~\ref{fig:dilpot}, significant deviations from the flat case
appear in the intermediate region where the derivatives of the
background fields are non-vanishing.

However, the potential remains  negative definite and does not allow for
the existence of bound states or quasi-bound states in the hyperstring
core. In that case, we can use the \wkb{} method to approximate the
solutions in the entire extra dimension since the turning points
where $\madim^2 = \Vtens{0}$ do not exist (see Fig.~\ref{fig:dilpot})
\begin{equation}
\tensrescal_{_\wkb}(\madim,z) \propto \dfrac{\exp\left(i \displaystyle \int^z
  \sqrt{\madim^2 - \Vtens{0}(u)} \, \ud u\right)}{\left[\madim^2 -
    \Vtens{0}(z) \right]^{1/4}}.
\end{equation}
After normalization at infinity one gets, from
Eq.~(\ref{eq:tensrescal}), a spectral density in the hyperstring core
similar to the one induced by the pure cylindrical waves
\begin{equation}
\left|\tensspec_{\madim}(0)\right|^2 = \ue^{-\phi(0)} \madim,
\end{equation}
and thus a $6D$ Newtonian potential, albeit with a weaker
gravitational coupling.

\begin{figure}
\begin{center}
\includegraphics[width=8.5cm]{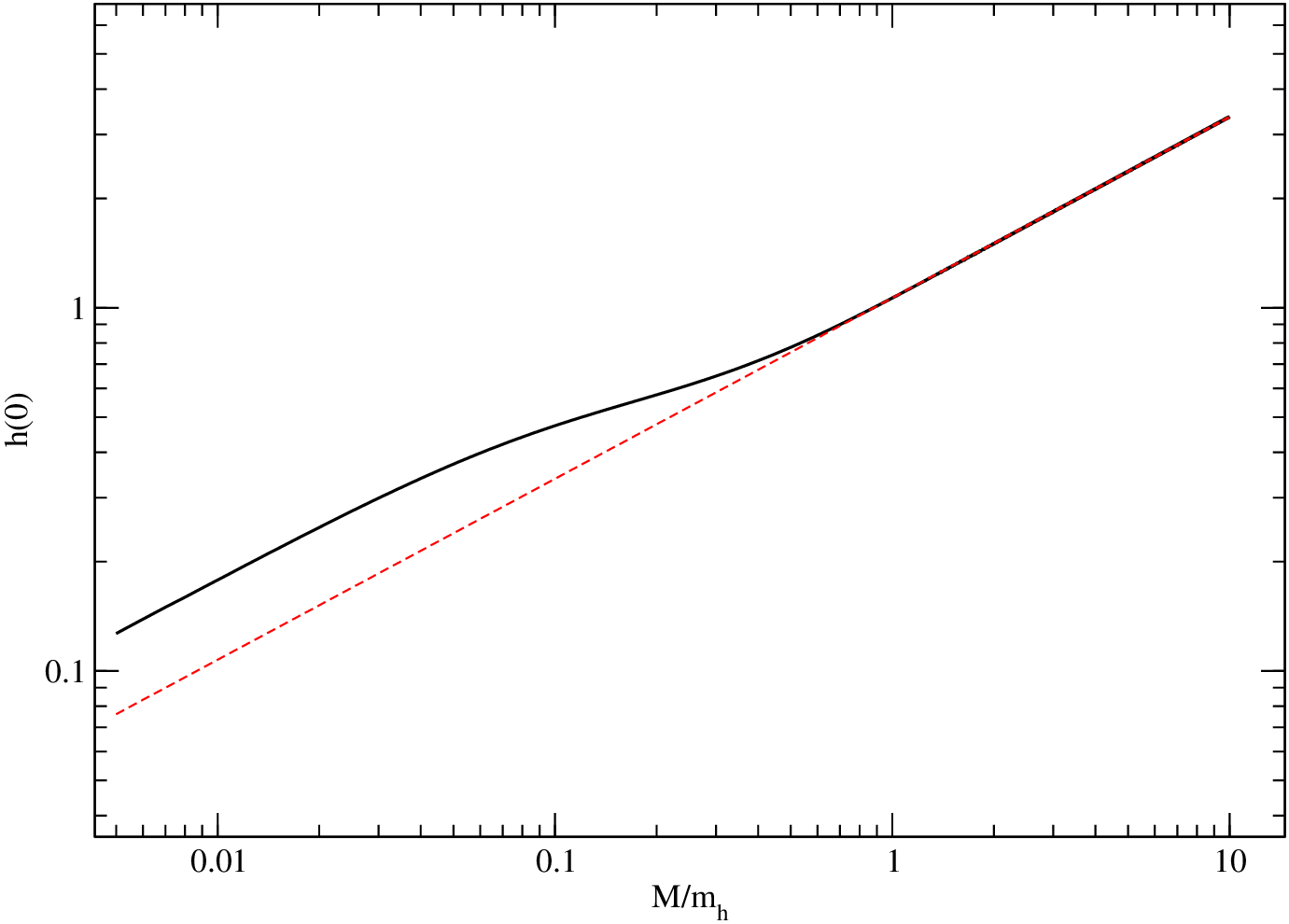}
\caption{Normalized values of $\tensspec(\madim,0)$ as function of
$\madim$ (solid line). For high and small values of $\madim$, one
recovers the $\sqrt{\madim}$ behaviour (dashed line). As expected from
the \wkb{} approximation, only small deviations from the $6D$ gravity
appear as long as the potential does not allow bound states.}
\label{fig:couplings}
\end{center}
\end{figure}

We have plotted in Fig.~\ref{fig:couplings} the values of
$\tensspec(\madim,0)$ obtained from numerical integrations of
Eq.~(\ref{eq:schrod}) for a wide range of masses $\madim$. This
confirms the \wkb{} result, with only slight deviations in the
intermediate range.

According to the previous discussions, this is physically expected as
long as the potential $\Vtens{0}$ remains negative definite: there are
no metastable $4D$ tensor waves inside the core that could induce
peaks in the spectral density. In the next section, we search for the
precise conditions under which the potential would in fact become
positive for a general codimension $\nextra=2$ defect and generic
dilaton potential.

\subsection{Confining Potentials and the Dominant Energy Condition}
\label{subsect:dec}
According to Eqs.~(\ref{eq:pot}) and (\ref{eq:superpot}), the
potential $\Vtens{0}$ driving the propagation of the non-winding
($p=0$) $4D$ tensor modes in the bulk reads
\begin{equation}
\label{eq:potstring}
2\Vtens{0} = \dfrac{1}{2} \left(\dfrac{3}{2} \sigma' +
\dfrac{\sqrgaa'}{\sqrgaa} + \phi' \right)^2
-\dfrac{\sqrgaa'^2}{\sqrgaa^2} + \dfrac{3}{2}\sigma'' +
\dfrac{\sqrgaa''}{\sqrgaa} + \phi''.
\end{equation}

{}From the topological defect stress tensor components (\ref{eq:tab}),
one can define the dimensionless matter energy $\calE$ and pressure
$\calP$ by
\begin{equation}
\label{eq:energy}
\begin{aligned}
\calE &\equiv \dfrac{1}{2} \left[ \dot{f}^2 + \dfrac{f^2
 Q^2}{\sqrgaa^2} + \dfrac{1}{4}\left(f^2-1\right)^2 +
 \dfrac{\dot{Q}^2}{\varepsilon \sqrgaa^2}\right], \\ \calP &\equiv
 \dfrac{1}{2} \left[ \dot{f}^2 - \dfrac{f^2 Q^2}{\sqrgaa^2} -
 \dfrac{1}{4}\left(f^2-1\right)^2 + \dfrac{\dot{Q}^2}{\varepsilon
 \sqrgaa^2} \right].
\end{aligned}
\end{equation}
The second order derivatives in Eq.~(\ref{eq:potstring}) can be
expressed in terms of first order derivatives by means of the dynamical
Einstein-Jordan equation (\ref{eq:Gmunu}). In terms of the conformal
coordinate $z$ and $\Uhat \equiv \exp{(\sigma)} U /\mhig^2$,
Eq.~(\ref{eq:Gmunu}) yields
\begin{equation}
\begin{aligned}
\dfrac{3}{2}\sigma'' & + \dfrac{\sqrgaa''}{\sqrgaa} + \phi'' =
-2\alpha \ue^{\sigma -\phi} \calE -\dfrac{1}{2} \Uhat -
\dfrac{3}{4} \sigma'^2 - \sigma' \dfrac{\sqrgaa'}{\sqrgaa}
\\ &-\dfrac{3}{2} \phi'^2 - \left(\sigma' + \dfrac{\sqrgaa'}{\sqrgaa}
\right) \phi'.
\end{aligned}
\end{equation}
Making use of the constraint equation (\ref{eq:Grr})
\begin{equation}
\begin{aligned}
\dfrac{3}{2} \sigma'^2 & - \dfrac{1}{2} \phi'^2 + 2 \sigma'
\dfrac{\sqrgaa'}{\sqrgaa} + \left(2 \sigma' +
\dfrac{\sqrgaa'}{\sqrgaa} \right) \phi' \\ & = 2 \alpha
\ue^{\sigma-\phi} \calP - \dfrac{1}{2} \Uhat,
\end{aligned}
\end{equation}
the potential $\Vtens{0}$ simplifies to
\begin{equation}
\label{eq:negpot}
2 \Vtens{0} = -2 \alpha \ue^{\sigma - \phi} \left(\calE - \calP
\right) - \dfrac{1}{2} \left(\dfrac{3}{2} \sigma' +
\dfrac{\sqrgaa'}{\sqrgaa} + \phi' \right)^2 - \Uhat,
\end{equation}
which, according to Eq.~(\ref{eq:energy}), is negative for positive
dilatonic potentials $U$. In fact it is clear in Eq.~(\ref{eq:negpot})
that, for a \emph{general} defect in codimension two, either
$\calE-\calP<0$ or $\Uhat<0$ are required for $\Vtens{0}$ to become
positive at some point. The former condition would violate the
Dominant Energy Condition (\dec{}) for the matter stress tensor, and
one would need ``special'' hyperstring forming fields that support
tachyonic propagation~\cite{Wald:1984rg}. It is beyond the scope of
this work to discuss more fundamental motivations and consequences of
negative dilatonic potentials~\cite{Linde:2001ae}, however we note
that the simplest example of this is achieved by choosing a negative
cosmological constant [see Eq.~(\ref{eq:action})]. In absence of a
dilaton, the six-dimensional spacetime generated by a hyperstring in
the presence of a negative cosmological constant has in fact already
been studied in
Refs.~\cite{Tinyakov:2001jt,Giovannini:2001hh,Peter:2003zg} and
explicitly realizes Randall-Sundrum gravity confinement. However, such
defect models require strong fine-tunings on the model parameters, and
stay in the realm of finite volume extra-dimensions. It is worth
pointing out that in this case, since the potential is asymptotically
positive due to the negative cosmological constant, there is a
normalizable zero mode (reminiscent of the trapped massless graviton
of the Randall-Sundrum model), which ensures that the potential
$\Vtens{0}$ does not break supersymmetry. Interestingly, one may also
expect to confine a discrete spectrum of massive gravitons.

In the next section, we explore the conditions under which there exist
quasi-localized $4D$ tensor waves on a classical defect in higher
codimensions, concentrating on the case in which the extra dimensional
space is isotropic.

\section{Classical Defects in Isotropic Extra Dimensions}

The action (\ref{eq:action}) can be generalized in a straightforward
way to a $\ntot$-dimensional spacetime and the Einstein-Jordan
equations remain given by Eqs.~(\ref{eq:tensor}) and
(\ref{eq:scalar}). The dynamics in the matter sector is driven by the
Lagrangian $\calL_\matter$ which should involve the required fields
and their interactions to allow the formation of a $4D$ topological
defect in $\ntot$ dimensions. In the simple case where the defect
forming fields do not break the $4D$ Poincar\'e invariance, \ie there
is no current flowing along the
defect~\cite{Witten:1984eb,Carter:1989xk,Peter:1992dw,Carter:1994hn},
the $4D$ components of the stress tensor verify
\begin{equation}
\label{eq:tmunuGN}
T_{\mu \nu} = -\calE g_{\mu\nu},
\end{equation}
where $\calE$ is a function of the transverse coordinates only. Like
the other components of the stress tensor, the profile of $\calE$ in
the extra dimensions, and consequently the behaviour of the metric
fluctuations $\delta g_{\mu\nu}$, depends of the underlying model
defined by $\calL_\matter$.

Here, we consider the generalization of Eq.~(\ref{eq:metric}) to more
than two isotropic extra
dimensions~\cite{Csaki:2000fc,Gherghetta:2000jf}
\begin{equation}
\label{eq:metricN}
\ud s^2=\ue^{\sigma(z)} \eta_{\mu \nu} \ud x^\mu \ud x^\nu +
\ue^{\sigma(z)} \ud z^2 + \sqrgaadim^2 (z) \ud \Omegan^2,
\end{equation}
where
\begin{equation}
\ud \Omegan^2 =  \sum_{i=1}^n \harm_i(\theta_{j<i}) \ud \theta_i^2,
\end{equation}
is the interval of a $n$ dimensional maximally symmetric
space. Depending on the curvature $\curv$ of this subspace, one may
choose the coordinates such as $\harm_1=1$ and
\begin{equation}
\harm_i(\theta_{j<i}) \equiv \harm_2(\theta_1)\prod_{j=2}^{i-1}
\sin^2(\theta_j),
\end{equation}
where $\harm_2$ is $\sin^2(\theta_2)$, $\theta_2^2$ or
$\sinh^2(\theta_2)$ for positive, null and vanishing curvature
$\curv$, respectively.

In this case, the Einstein-Jordan equations (\ref{eq:tensor}) read
\begin{equation}
\label{eq:tensormunuN}
\begin{aligned}
\frac{3}{2} \sigma'' + n \dfrac{\sqrgaadim''}{\sqrgaadim} + \phi'' & =
-\kappan^2 \ue^{\sigma- \phi} \calE - \dfrac{1}{2} \ue^{\sigma} U +
\dfrac{\ue^\sigma}{\sqrgaadim^2} \dfrac{n(n-1)}{2}\curv \\
&-\dfrac{3}{4} \sigma'^2 - \dfrac{n(n-1)}{2}
\dfrac{\sqrgaadim'^2}{\sqrgaadim^2}  - \dfrac{3}{2} \phi'^2 \\ &- n \sigma'
\dfrac{\sqrgaadim'}{\sqrgaadim} - \left(\sigma' + n
\dfrac{\sqrgaa'}{\sqrgaa} \right) \phi',
\end{aligned}
\end{equation}
for the $(\mu\nu)$ components, while the constraint equation $(zz)$ is
\begin{equation}
\label{eq:tensorzzN}
\begin{aligned}
\dfrac{3}{2} \sigma'^2 & - \dfrac{1}{2} \phi'^2 + \dfrac{n(n-1)}{2}
\dfrac{\sqrgaadim'^2}{\sqrgaadim^2} + 2n \sigma'
\dfrac{\sqrgaadim'}{\sqrgaadim} - \dfrac{\ue^\sigma}{\sqrgaadim^2}
\dfrac{n(n-1)}{2}\curv\\ &+ \left(2 \sigma' + n
\dfrac{\sqrgaa'}{\sqrgaa} \right) \phi' = \kappan^2 \ue^{\sigma-\phi}
\calP - \dfrac{1}{2} \ue^{\sigma} U,
\end{aligned}
\end{equation}
with $\calP$ the pressure along the $z$ extra dimension. The tensor
perturbations $\delta g_{\mu\nu}=\exp(\sigma) \tens_{\mu\nu}$ around
this metric end up being solutions of
\begin{equation}
\label{eq:tensmotionN}
\tens''_{\mu \nu} + \left(\dfrac{3}{2} \sigma' + n
\dfrac{\sqrgaadim'}{\sqrgaadim} + \phi' \right) \tens'_{\mu \nu} +
\Boxadim \tens_{\mu \nu} +
\dfrac{\ue^{\sigma}}{\sqrgaadim^2}\angmom_n\tens_{\mu \nu} = 0,
\end{equation}
where $\angmom_n$ is a generalized ``angular'' differential operator,
\eg for $\curv=1$ one recovers the hyperspherical Laplacian
\begin{equation}
\angmom_n = \sum_{i=1}^n \dfrac{1}{\harm_i}\left(\partial_{\theta_i}^2 +
  \dfrac{n-i}{\tan\theta_i} \partial_{\theta_i} \right).
\end{equation}
For the zero angular momentum modes, \ie those which are non-vanishing
on the brane for symmetry reasons, one recovers the Schr\"odinger
equation (\ref{eq:schrod}) for the rescaled quantity
\begin{equation}
\tensrescal_{\alpha \beta} = \ue^{3\sigma/4 + \phi/2} \sqrgaadim^{n/2} \,
\tensspec_{\alpha \beta}.
\end{equation}
Using the above Einstein-Jordan equations, as described in
Sect.~\ref{subsect:dec}, the potential is found to be
\begin{equation}
\begin{aligned}
\label{eq:pottensN}
2\Vtens{0} & = -\kappan^2 \ue^{\sigma-\phi} \left(\calE - \calP\right)
- \ue^{\sigma} U + \dfrac{\ue^{\sigma}}{\sqrgaadim^2}n(n-1) \curv \\
&- \dfrac{1}{2} \left(\dfrac{3}{2} \sigma' + n
\dfrac{\sqrgaadim'}{\sqrgaadim} + \phi' \right)^2.
\end{aligned}
\end{equation}

In contrast to the hyperstring potential studied before [see
Eq.~(\ref{eq:negpot})], it appears that for positive curvature
$\curv=1$ the potential may now take positive values, whereas for
$\curv=0,-1$, one would still need to violate the \dec{} to allow for
quasi-localized states .This result is not really surprising, as often
in General Relativity, a standard way to mimic matter that violates
positive energy conditions is to consider positive curvature space
(\eg see Refs.~\cite{Kaloper:2000jb, Carroll:2001ih, Peter:2001fy,
Martin:2003sf}). {}From Eq.~(\ref{eq:metricN}), note that $\curv$ is
the curvature of $n=\nextra-1$ extra dimensional subspace and,
according to the asymptotic behaviour of $\sqrgaadim(z)$, the volume
of the total $\nextra$ extra dimensions may be infinite even for
$\curv=1$.

Although we have not presented the existence of solitonic solutions
compatible with the metric (\ref{eq:metricN}), the condition $k=1$ may
be generically fulfilled. For example the case of $\nextra=3$ extra
dimensions with a positive curvature $\curv$ may be realized by a
hypermonopole~\cite{Dvali:2000ty, Csaki:2000fc,Roessl:2002rv}. In such
a framework one may indeed expect to find quasi-localized tensor
modes in the core without violating the \dec{}~\cite{us:2004}.

\section{Conclusion}
\label{sec:conc}
We have discussed in this paper the conditions under which one can
realize quasi-localized gravity using underlying topological defect
models. For defects with codimension $\nextra=2$ the existence of
metastable tensor modes requires violation of the Dominant Energy
Condition by the defect matter. For codimension $\nextra>2$ we found
that this requirement may be relaxed for appropriate defect matter. In
this case the background solution can, at least in principle, support
metastable gravity. The explicit construction of $\nextra>2$
codimensional sigma models that quasi-localize gravity is subject to
future work~\cite{us:2004}.

\acknowledgments

We thank Gregory Gabadadze, Patrick Peter, Massimo Porrati and
Jean-Philippe Uzan for enlightening discussions and useful
comments. J-W.~R. thanks the I.A.P., where part of this work was done,
for their warm hospitality.

\appendix

\section{Relaxation method}
\label{sect:appnum}

To solve the set of differential equations (\ref{eq:Gmunu}),
(\ref{eq:Gaa}), (\ref{eq:dil}), (\ref{eq:kg}) and (\ref{eq:max}), we
have used a relaxation method~\cite{Adler:1983zh}. In terms of
dimensionless fields and parameters, the action (\ref{eq:action}) can
be recast into
\begin{equation}
\label{eq:actionbypart}
\begin{aligned}
-\dfrac{2 \alpha}{\pi \eta^2} S&=\int \ue^{\phi} \Big[-3
 \dot{\expsig}^2 \sqrgaa - 4 \expsig \dot{\expsig} \dot{\sqrgaa} _ +
 \expsig^2 \sqrgaa \left(\dot{\phi}^2 + \beta \phi^2 \right) \\ & -
 \dot{\phi} \left(4 \expsig \dot{\expsig} \sqrgaa + 2 \dot{\sqrgaa}
 \expsig^2 \right) \Big] \ud \radim  + 2 \alpha \int \expsig^2
 \sqrgaa \Bigg[ \dot{f}^2 \\ &+ \dfrac{f^2 Q^2}{\sqrgaa^2} + \dfrac{1}{4}
 \left(f^2-1 \right)^2 + \dfrac{\dot{Q}^2}{\varepsilon \sqrgaa^2}
 \Bigg] \ud \radim,
\end{aligned}
\end{equation}
with $\expsig=\ue^{\sigma}$, and where integrations by parts have been
used to keep only first order derivatives in the metric factors. After
discretization of the radial coordinates $\radim$, the discrete action
reads $\Sdis=\Sgdis + \Smdis$ with
\begin{widetext}
\begin{equation}
\label{eq:Sgdis}
\begin{aligned}
-\dfrac{2\alpha}{\pi \eta^2} \Sgdis &= \sum_{i} \ue^{\phi_i} \Bigg[ -
  3 \sqrgaa_i
  \dfrac{\left(\expsig_{i+1/2}-\expsig_{i-1/2}\right)^2}{h} -
  4\expsig_i \dfrac{\left(\expsig_{i+1/2} -\expsig_{i-1/2}
    \right)\left( \sqrgaa_{i+1/2} - \sqrgaa_{i-1/2} \right)}{h} +
  \expsig_i^2 \sqrgaa_i \dfrac{\left(\phi_{i+1/2} - \phi_{i-1/2}
    \right)^2}{h} \\ & - 4\expsig_i \sqrgaa_i
  \dfrac{\left(\expsig_{i+1/2} -\expsig_{i-1/2} \right)\left( \phi_{i+1/2} -
    \phi_{i-1/2} \right)}{h} - 2\expsig_i^2
  \dfrac{\left(\sqrgaa_{i+1/2} -\sqrgaa_{i-1/2} \right)\left( \phi_{i+1/2} -
    \phi_{i-1/2} \right)}{h} + h \beta \expsig_i^2 \sqrgaa_i
  \phi_i^2\Bigg],
\end{aligned}
\end{equation}
in the gravity sector, and
\begin{equation}
\label{eq:Smdis}
\begin{aligned}
-\dfrac{\Smdis}{\pi \eta^2} &= \sum_{i} \Bigg[ \expsig_i^2 \sqrgaa_i
    \dfrac{\left( f_{i+1/2} - f_{i-1/2} \right)^2}{h} + h \expsig_i^2
    \dfrac{f_i^2 Q_i^2}{\sqrgaa_i} + h \dfrac{\expsig_i^2
    \sqrgaa_i}{4} \left(f_i^2 -1 \right)^2 +
    \dfrac{\expsig_i^2}{\varepsilon \sqrgaa_i}
    \dfrac{\left(Q_{i+1/2}-Q_{i-1/2} \right)^2}{h} \Bigg],
\end{aligned}
\end{equation}
\end{widetext}
for the matter sector. The grid resolution is given by $h=\radim_{i+1} -
\radim_{i}$ where the index $i$ indicates that the fields have to be
evaluated at the discrete points $\radim_i$. The discrete derivatives
have been expressed in their centered form, \eg
\begin{equation}
\dot{f}_i = \dfrac{f_{i+1/2} - f_{i-1/2}}{h} + \mathcal{O}(h^3),
\end{equation}
where $f_{i+1/2}$ is evaluated on a half step shifted mesh. By
differentiating the discrete action $\Sdis$ with respect to
the discrete fields $\expsig_i$, $\sqrgaa_i$, $\phi_i$, $f_i$ and
$Q_i$, one gets the finite difference equations corresponding to
Eqs.~(\ref{eq:Gmunu}), (\ref{eq:Gaa}), (\ref{eq:dil}), (\ref{eq:kg})
and (\ref{eq:max}), respectively. {}From an initial guess of all the
discrete fields $\vec{\calF}_i^0$ on the grid $\radim_i$, the
solutions of the finite difference equations are obtained by a
successive over-relaxation method. Here $\vec{\calF}_i$ designs a five
dimensional vector whose components are the discrete fields
$\expsig_i$, $\sqrgaa_i$, $\phi_i$, $f_i$ and $Q_i$. At step $p+1$,
the fields are updated by a Newton's method to reduce the error with
respect to the true solution:
\begin{equation}
\label{eq:sor}
\vec{\calF}_i^{p+1} =\vec{\calF}_i^{p} + s \,\delta
\vec{\calF}_i^p,
\end{equation}
where $\delta \vec{\calF}_i^p$ is computed to solve
\begin{equation}
\vec{\calE}\negthinspace\left(\vec{\calF}_i^{p+1}\right) \simeq
\vec{\calE}\negthinspace\left(\vec{\calF}_i^{p} \right) + \vec{\nabla}
\vec{\calE}
\negthinspace\left(\delta \vec{\calF}_i^p \right) =0,
\end{equation}
$\vec{\calE}$ being the finite difference equations. In
Eq.~(\ref{eq:sor}), $s$ is the over-relaxation factor. The boundary
conditions are part of the finite difference equations $\vec{\calE}$
since they appear as the constraints which have to be satisfied by
$\vec{\calF}_1^p$ and $\vec{\calF}_N^p$, where $N$ is the total number
of points of the $\radim_i$ mesh. The iterative process is stopped
when the discretized action $\Sdis$ given by Eqs.~(\ref{eq:Sgdis}) and
(\ref{eq:Smdis}) remains stationary at the machine
precision. Moreover, we have also verified that the relaxed solutions
satisfy the constraint equation (\ref{eq:Grr}). The relaxed fields
have been plotted in Fig.~\ref{fig:dilvort} for an assumed generic set
of parameters $\alpha$, $\beta$ and $\epsilon$.

\section{Perturbed quantities}
\label{sect:apppert}
For the metric (\ref{eq:metricN}), the only non-vanishing tensor
perturbations are
\begin{equation}
\label{eq:pertg}
\delta g_{\mu \nu} = \ue^{\sigma} \tens_{\mu \nu}, \quad \delta g^{\mu
\nu} = - \ue^{-\sigma} \eta^{\mu \alpha} \eta^{\nu \beta} \tens_{\alpha
\beta},
\end{equation}
from which the perturbed Christoffel symbols read
\begin{equation}
\begin{aligned}
\delta \Gamma^{\alpha}_{\mu \nu} & = \dfrac{1}{2} \eta^{\alpha \beta}
\left( \tens_{\beta \mu,\nu} + \tens_{\beta \nu,\mu} - \tens_{\mu \nu,
\beta} \right), \\ \delta \Gamma^{\alpha}_{\mu z} &= \dfrac{1}{2}
\eta^{\alpha \beta} \tens'_{\mu \beta}, \\ \delta \Gamma^{\alpha}_{\mu
\theta_i} &= \dfrac{1}{2} \eta^{\alpha \beta} \partial_{\theta_i}
\tens_{\mu \beta}, \\ \delta \Gamma^{z}_{\mu \nu} &= -\dfrac{1}{2}
\left(\tens'_{\mu \nu} + \sigma' \tens_{\mu \nu} \right),\\ \delta
\Gamma^{\theta_i}_{\mu \nu} &= -\dfrac{1}{2}
\dfrac{\ue^{\sigma}}{\sqrgaadim^2} \dfrac{\partial_{\theta_i}
\tens_{\mu \nu}}{\harm_i}.
\end{aligned}
\end{equation}
{}From the perturbed Ricci tensor
\begin{equation}
\begin{aligned}
\delta R_{\si{BD}} &= - \delta\Gamma^{\si{A}}_{\si{BA,D}} +
\delta\Gamma^{\si{A}}_{\si{BD,A}} + \Gamma^{\si{H}}_{\si{BD}} \delta
\Gamma^{\si{A}}_{\si{AH}} - \Gamma^{\si{A}}_{\si{DH}} \delta
\Gamma^{\si{H}}_{\si{BA}} \\ & - \Gamma^{\si{H}}_{\si{BA}} \delta
\Gamma^{\si{A}}_{\si{DH}} + \Gamma^{\si{A}}_{\si{AH}} \delta
\Gamma^{\si{H}}_{\si{BD}},
\end{aligned}
\end{equation}
the non-vanishing components of the perturbed Einstein tensor read
\begin{equation}
\label{eq:pertgmunu}
\begin{aligned}
\delta G_{\mu \nu} &= - \dfrac{1}{2} \left\{ \tens_{\mu \nu}'' +
    \left(\dfrac{3}{2} \sigma' + n\dfrac{\sqrgaadim'}{\sqrgaadim}
    \right) \tens_{\mu \nu}' + \Box\tens_{\mu \nu} \right. \\ &+
    \left.  \dfrac{\ue^{\sigma}}{\sqrgaadim^2}\angmom_n \tens_{\mu
    \nu} - \left[3 \sigma'' + 2n \dfrac{\sqrgaadim''}{\sqrgaadim} +
    \dfrac{3}{2} \sigma'^2 \right.\right.\\ & \left.\left. +
    n(n-1)\dfrac{\sqrgaadim'^2}{\sqrgaadim^2} + 2n \sigma'
    \dfrac{\sqrgaadim'}{\sqrgaadim} -
    \dfrac{\ue^{\sigma}}{\sqrgaadim^2} n(n-1) \curv \right] \tens_{\mu
    \nu} \right\}.
\end{aligned}
\end{equation}
{}From Eqs.~(\ref{eq:tensor}) and (\ref{eq:pertg}), the perturbed
dilatonic source terms, \ie other than the matter stress tensor, are
\begin{equation}
\label{eq:pertdil}
\begin{aligned}
\delta \calD_{\mu \nu} &= \dfrac{1}{2} \phi' \tens_{\mu \nu}' -
 \left[\phi'' + \dfrac{3}{2} \phi'^2 +
  \left(\sigma' + n\dfrac{\sqrgaadim'}{\sqrgaadim} \right)
  \phi'\right. \\ & + \left. \dfrac{1}{2} U(\phi) \right] \tens_{\mu \nu},
\end{aligned}
\end{equation}
whereas $\delta T_{\mu \nu}$ is readily obtained from
Eq.~(\ref{eq:tmunuGN}), or Eq.~(\ref{eq:tab}) for the Abelian Higgs
vortex. {}From Eqs.~(\ref{eq:pertgmunu}), and (\ref{eq:pertdil}), one
obtains Eqs.~(\ref{eq:tensmotion}) and (\ref{eq:tensmotionN}), up to
the background Einstein-Jordan equation (\ref{eq:tensor}).

\bibliography{bibdilastring}

\end{document}